\documentclass[11pt]{article}
\usepackage{amsfonts}
\usepackage{amssymb}
\usepackage{amsmath}
\usepackage{a4wide}
\usepackage{graphicx}

\newtheorem{definition}{Definition}

\makeatletter
\def \@removefromreset#1#2{\let \@tempb \@elt
     \def \@tempa#1{@&#1}\expandafter \let \csname @*#1*\endcsname \@tempa
     \def \@elt##1{\expandafter \ifx \csname @*##1*\endcsname \@tempa \else
    \noexpand \@elt{##1}\fi}     \expandafter \edef \csname cl@#2\endcsname{\csname cl@#2\endcsname}     \let \@elt \@tempb
     \expandafter \let \csname @*#1*\endcsname \@undefined}

\@removefromreset{equation}{section}

\@removefromreset{theorem}{section}
\makeatother
\input{tcilatex}
\begin{document}

\date{}
\title{On the existence of a local quasi hidden variable (LqHV) model for
each $N$-qudit state and the maximal quantum violation of Bell inequalities}
\author{Elena R. Loubenets \\
Moscow Institute of Electronics and Mathematics, \\
National Research University HSE, Moscow, 123458, Russia}
\maketitle

\begin{abstract}
We specify the local quasi hidden variable (LqHV) model reproducing the
probabilistic description of all $N$-partite joint von Neumann measurements
on an $N$-qudit state. Via this local probability model, we derive a new
upper bound on the maximal violation by an $N$-qudit state of $N$-partite
Bell inequalities of any type (either on correlation functions or on joint
probabilities) for $S$ observables per site. This new upper bound not only
improves for all $N,$ $S$ and $d$ the corresponding results available for
general Bell inequalities in the literature but also, for the $N$-qubit case
with two observables per site, reduces exactly to the attainable upper bound
known for quantum violations of $2\times \cdots \times 2$-setting
correlation Bell inequalities in a dichotomic case.\medskip

\end{abstract}

\section{Introduction}

In 1935, Einstein, Podolsky and Rosen (EPR) argued \cite{2} that \emph{%
locality} of measurements performed by two parties on perfectly correlated
quantum events implies the "simultaneous reality - and thus definite values"%
\footnote{%
See \cite{2}, page 778.} of physical quantities described by noncommuting
quantum observables. This EPR argument, contradicting the quantum formalism
and known as the EPR paradox, seemed to imply a possibility of a \emph{%
hidden variable} account of quantum measurements.

Analyzing this possibility in 1964-1966, Bell explicitly constructed \cite{3}
the hidden variable (HV) model reproducing the statistical properties of all
qubit observables. Considering, however, spin measurements of two parties on
the two-qubit singlet state, Bell proved \cite{4} that any local hidden
variable (LHV) description of these bipartite joint spin measurements on
perfectly correlated quantum events disagrees with the statistical
predictions of quantum theory. In view of his mathematical results in \cite%
{3, 4}, Bell argued \cite{3, 5} that the EPR paradox should be resolved
specifically due to violation of \emph{locality} under multipartite quantum
measurements and that "... non-locality is deeply rooted in quantum
mechanics itself and will persist in any completion".

However, as we stressed in Sec. 3 of \cite{10}, though both specifications
of locality, one by EPR in \cite{2} and another by Bell in \cite{4},
correspond to the manifestation of the physical principle of\emph{\ local
action }under multipartite nonsignaling measurements, but -- the EPR
locality, described in \cite{2} as "without in any way disturbing" systems
and measurements at other sites, is a general concept, not in any way
associated with the use of some specific mathematical formalism, whereas
Bell's locality (as it is formulated in \cite{4}) constitutes the
manifestation of locality specifically in the HV frame. As a result, Bell's
locality implies the EPR locality, but the converse is not true -- the EPR
locality \emph{does not need} to imply Bell's locality so that the proved by
Bell non-existence of a local HV (LHV) model for the singlet state does not
point to resolution of the EPR paradox via violation of the EPR locality.
For details, see Sec. 3 in \cite{10}.

Nowadays, there is still no a unique conceptual view\footnote{%
See, for example, discussions in \cite{RL, KH, KH1, Gis}.} on Bell's concept
of quantum nonlocality. However, it is clear that this concept does not mean
propagation of interaction faster than light and that it is not equivalent
to the concept of quantum entanglement. Moreover, in quantum information, a
nonlocal multipartite quantum state is defined purely mathematically - via
violation by this state of some Bell inequality\footnote{%
The general frame for multipartite Bell inequalities for arbitrary numbers
of settings and outcomes per site was introduced in \cite{11}.} or,
equivalently, via non-existence for each quantum correlation scenario on
this state of a local HV (LHV)\ model.

Note that, from the mathematical point of view, the violation result of Bell
in \cite{4} (known in the physical literature as Bell's theorem) can be
conditioned by either of at least two mathematical alternatives: (i) the
dependence of a random variable at one site not only on an observable
measured at this site but also on measurement settings and outcomes at the
other sites; (ii) non-positivity of a scalar measure $\nu $ modelling the
singlet state. From the physical point of view, a choice between these two
mathematical alternatives corresponds to a choice between (i)\textbf{\ }%
\emph{nonlocality }and (ii) \emph{nonclassicality.} The latter corresponds
to violation of "classical realism" embedded into HV models via probability
measures.

Moreover, as we proved by theorem 2 in \cite{12}, for the probabilistic
description of every quantum correlation scenario, the second alternative
does always work. Namely, each quantum correlation scenario admits \emph{a
local quasi hidden variable (LqHV) model} -- a new general local probability
model which we introduced in \cite{12, 13} and where all averages and
product expectations of a correlation scenario are reproduced via \emph{local%
} random variables on a measure space\footnote{%
Here, $\Omega $ is a set, $\mathcal{F}_{\Omega }$ is an algebra of subsets
of $\Omega $ and $\mu $ is a measure on $\mathcal{F}_{\Omega }$.} $(\Omega ,%
\mathcal{F}_{\Omega },\mu )$ but, in this triple, a measure $\mu $ is
real-valued and does not need to be positive\footnote{%
For two canonically conjugate quantum observables, the representation of
their averages via the real-valued measure, \emph{the Wigner quasi
probability distribution}, was first introduced by Wigner \cite{Wig}.}. Note
that, under the LqHV modelling of a quantum correlation scenario there are
no negative probabilities\footnote{%
The introduction of negative probabilities into the quantum formalism was
first suggested by Dirac \cite{Dir} and analysed further by Feynman \cite%
{Fey}.} -- though a measure $\mu $ can have negative values, all scenario
joint probabilities are expressed only via nonnegative values of this
measure.

Furthermore, from our recent proof \cite{new} of the existence of \emph{a
context-invariant qHV model} \emph{for all quantum observables and states}
on an arbitrary Hilbert space it follows (see proposition 3 in \cite{new})
that \emph{every }$N$\emph{-partite quantum state admits a LqHV model, }that
is, a single LqHV model for all $N$-partite joint von Neumann measurements
on this state.\emph{\ }This\emph{\ }new notion is similar by its form to
Werner's \cite{werner} notion of a LHV model for an $N$-partite quantum
state but with replacement of a positive measure in the representation for $%
N $-partite joint probabilities by a real-valued one. Recall that an
arbitrary entangled quantum state does not need to admit a LHV model.

All these new results in \cite{12, 13, new} indicate that, from the point of
view of mathematical modelling, the choice of the second above alternative,
resulting in the LqHV representations for joint probabilities and product
expectations, which are always valid, is much more justified than Bell's
choice of the first alternative leading to the conjecture \cite{3, 5} on
quantum nonlocality.

Moreover, it is specifically this new type of probabilistic modelling, \emph{%
the LqHV modelling,} that allowed us to derive \cite{12, found} the upper
bounds on quantum violations of \emph{general}\footnote{%
That is, Bell inequalities of any possible type - either on correlation
functions, full or resticted, or on joint probabilities or on both. For the
definition of a Bell inequality, its general form and specific examples, see 
\cite{11}.} Bell inequalities which essentially improve the corresponding
results available \cite{kaplan, carlos, carlos2, carlos3} via other
mathematical frames, in \ particular, via the operator space theory in \cite%
{carlos, carlos2, carlos3}.

In the present paper, we analyse further the computational capabilities of
the LqHV modelling. Via the LqHV frame, we find a new upper bound on
violations by an $N$-qudit state of general Bell inequalities for $S$
settings per site. This new upper bound incorporates and improves our
results in \cite{12, found}. In the $N$-qubit case with two observables per
site, it reduces exactly to the attainable upper bound known \cite{wolf} for
quantum violations of correlation $2\times \cdots \times 2$-setting Bell
inequalities in a dichotomic case.

The present paper is organized as follows.

In Sec. 2, we present the specific LqHV model for an $N$-qudit state.

In Sec. 3, we express in the LqHV terms the maximal violation by an $N$%
-qudit state of general Bell inequalities and, using the LqHV model
introduced in Sec. 2, we find a new upper bound on the maximal violation by
an $N$-qudit state of general Bell inequalities for $S$ settings per site.

In Sec. 4, we compare our new general upper bound with the upper bounds
available now in the literature.

In Sec. 5, we formulate the main results and stress the advantages of the
LqHV modelling.

\section{The LqHV model for an N-qudit state}

Denote by $\mathfrak{X}_{d^{N}}$ the set of all $N$-qudit observables $X$ on 
$(\mathbb{C}^{d})^{\otimes N}$ and by $\Lambda $ the set of all real-valued
functions 
\begin{equation}
\lambda :\mathfrak{X}_{d^{N}}\rightarrow \dbigcup \limits_{X\in \mathfrak{X}%
_{d^{N}}}\mathrm{sp}X
\end{equation}%
with values $\lambda (X)\equiv \lambda _{X}$ in the spectrum $\mathrm{sp}X$
of the corresponding observable $X$. Let $\pi _{(X_{1},...,X_{m})}:$ $%
\Lambda \rightarrow \mathrm{sp}X_{1}\times \cdots \times \mathrm{sp}X_{m}$
be the canonical projection on $\Lambda :$ 
\begin{eqnarray}
\pi _{(X_{1},...,X_{m})}(\lambda ) &=&\left( \pi _{X_{1}}(\lambda ),...,\pi
_{X_{m}}(\lambda )\right) ,\text{ \  \ }m\in \mathbb{N},  \label{1} \\
\pi _{X}(\lambda ) &=&\lambda _{X}\in \mathrm{sp}X,  \notag
\end{eqnarray}%
and $\mathcal{A}_{\Lambda }$ be the algebra of all cylindrical subsets of $%
\Lambda $ of the form 
\begin{eqnarray}
\pi _{(X_{1},...,X_{m})}^{-1}(F) &=&\{ \lambda \in \Lambda \mid (\pi
_{X_{1}}(\lambda ),...,\pi _{X_{m}}(\lambda ))\in F\},  \label{2} \\
F &\subseteq &\mathrm{sp}X_{1}\times \cdots \times \mathrm{sp}X_{m},  \notag
\end{eqnarray}%
for all collections $\{X_{1},...,X_{m}\} \subset \mathfrak{X}_{d^{N}},$ $%
m\in \mathbb{N},$ of $N$-qudit observables.

By Proposition 1 and relations (20), (35) in \cite{new}, to every $N$-qudit
state $\rho _{d,N}$ on $(\mathbb{C}^{d})^{\otimes N}$, there corresponds the
unique normalized finitely additive real-valued measure $\mu _{\rho _{d,N}}$
on $\mathcal{A}_{\Lambda },$ defined via the relation%
\begin{eqnarray}
\mu _{\rho _{d,N}}(\pi _{(X_{1},...,X_{m})}^{-1}(F)) &=&\frac{1}{m!}\dsum
\limits_{(x_{1},...,x_{m})\in F}\mathrm{tr}[\rho _{d,N}\{ \mathrm{P}%
_{X_{1}}(x_{1})\cdot \ldots \cdot \mathrm{P}_{X_{m}}(x_{m})\}_{\mathrm{sym}%
}],  \label{3} \\
F &\subseteq &\mathrm{sp}X_{1}\times \cdots \times \mathrm{sp}X_{m},  \notag
\end{eqnarray}%
for all sets $\pi _{(X_{1},...,X_{m})}^{-1}(F)\in \mathcal{A}_{\Lambda }$
and all finite collections $\{X_{1},...,X_{m}\} \subset \mathfrak{X}_{d^{N}}$
of $N$-qudit observables. Here, $\mathrm{P}_{X}(\cdot )$ is the spectral
measure of an $N$-qudit observable $X$ and the notation $\{Z_{1}\cdot \ldots
\cdot Z_{m}\}_{sym}$ means the operator sum corresponding to the
symmetrization of the operator product $Z_{1}\cdot \ldots \cdot Z_{m}$ with
respect to all permutations of its factors. From (\ref{3}) and the spectral
theorem it follows 
\begin{equation}
\frac{1}{m!}\mathrm{tr}[\rho _{d,N}\{X_{1}\cdot \ldots \cdot X_{m})\}_{%
\mathrm{sym}}]=\dint \limits_{\Lambda }\pi _{X_{1}}(\lambda )\cdot \ldots
\cdot \pi _{X_{m}}(\lambda )\mu _{\rho _{d,N}}(\mathrm{d}\lambda ).
\label{4}
\end{equation}%
\medskip

For an $N$-qudit state $\rho _{d,N}$ and arbitrary $N$-qudit observables of
the form 
\begin{eqnarray}
\widetilde{X}_{n} &=&\mathbb{I}_{(\mathbb{C}^{d})^{\otimes (n-1)}}\mathbb{%
\otimes }X_{n}\otimes \mathbb{I}_{(\mathbb{C}^{d})^{\otimes (N-n)}},
\label{5} \\
X_{n} &\in &\mathfrak{X}_{d},\text{ \  \ }n=1,...,N,  \notag
\end{eqnarray}%
relations (\ref{3}), (\ref{4}) imply the representations\footnote{%
Here, $\chi _{A}(\lambda )$ is the indicator function of a subset $%
A\subseteq \Lambda ,$ i. e. $\chi _{A}(\lambda )=1,$ if $\lambda \in A,$ and 
$\chi _{A}(\lambda )=0,$ if $\lambda \notin A$.} 
\begin{eqnarray}
&&\mathrm{tr}[\rho _{d,N}\{ \mathrm{P}_{X_{1}}(B_{1})\otimes \cdots \otimes 
\mathrm{P}_{X_{N}}(B_{N})\}]  \label{6} \\
&=&\dint \limits_{\Lambda }\chi _{\pi _{\widetilde{X}_{1}}^{-1}(B_{1})}(%
\lambda )\cdot \ldots \cdot \chi _{\pi _{\widetilde{X}_{N}}^{-1}(B_{N})}(%
\lambda )\text{ }\mu _{\rho _{d,N}}(\mathrm{d}\lambda ),\text{ \  \ }%
B_{n}\subseteq \mathrm{sp}X_{n},  \notag
\end{eqnarray}%
and%
\begin{equation}
\mathrm{tr}[\rho _{d,N}\{X_{1}\otimes \cdots \otimes X_{N}\}]=\dint
\limits_{\Lambda }\pi _{\widetilde{X}_{1}}(\lambda )\cdot \ldots \cdot \pi _{%
\widetilde{X}_{N}}(\lambda )\mu _{\rho _{d,N}}(\mathrm{d}\lambda ),
\label{7}
\end{equation}%
specified in terms of the measure space $(\Lambda ,\mathcal{A}_{\Lambda
},\mu _{\rho _{d,N}})$ where the normalized measure $\mu _{\rho _{d,N}}$ is
real-valued and the random variables $\pi _{\widetilde{X}_{n}}(\lambda ),$ $%
n=1,...,N,$ are \emph{local }in the sense that each of them depends only on
the corresponding observable $X_{n}$ at $n$-th site.

Consider the following general notion introduced in \cite{12, new}.

\begin{definition}
An N-partite quantum state $\rho $ admits a local qHV (LqHV) model if, for
all observables $X_{n}$ on each $n$-th site, all N-partite joint von Neumann
probabilities 
\begin{equation}
\mathrm{tr}[\rho \{ \mathrm{P}_{X_{1}}(B_{1})\otimes \cdots \otimes \mathrm{P%
}_{X_{N}}(B_{N})\}],\text{ \  \  \ }B_{n}\subseteq \mathrm{sp}X_{n},
\end{equation}%
admit the representation 
\begin{equation}
\mathrm{tr}[\rho \{ \mathrm{P}_{X_{1}}(B_{1})\otimes \cdots \otimes \mathrm{P%
}_{X_{N}}(B_{N})\}]=\dint \limits_{\Omega }P_{X_{1}}(B_{1};\omega )\cdot
\ldots \cdot P_{X_{N}}(B_{N};\omega )\nu _{\rho }(\text{\textrm{d}}\omega )
\label{8}
\end{equation}%
in terms of a single measure space $(\Omega ,\mathcal{F}_{\Omega },\nu
_{\rho })$ with a normalized real-valued measure $\nu _{\rho }$ and
conditional probability distributions $P_{Y_{n}}(\cdot $ $;\omega ),$ $%
n=1,...,N,$ each depending only on the corresponding observable $X_{n}$ at $%
n $-th site.
\end{definition}

By this definition, \emph{every }$N$\emph{-qudit state }$\rho _{d,N}$ \emph{%
admits the LqHV model (\ref{6}), }specified by relations (\ref{1}) - (\ref{3}%
).

\section{Quantum violations of general Bell inequalities}

In this section, we use the LqHV model (\ref{6}) for finding a new upper
bound on the maximal violation by an $N$-qudit state of \emph{general}%
\footnote{%
See footnote 7.} Bell inequalities,.

Consider a correlation scenario, performed on an $N$-qudit state $\rho
_{d,N} $\ and with $S$ qudit observables $X_{n}^{(s)},$ $s=1,...,S,$
measured projectively at each $n$-th site. By restricting the measure $\mu
_{\rho _{d,N}},$ given on the algebra $\mathcal{A}_{\Lambda }$ of subsets of 
$\Lambda $ by relation (\ref{3}), to the subalgebra of all cylindrical
subsets of the form 
\begin{equation}
\pi _{(\widetilde{X}_{1}^{(1)},...,\widetilde{X}_{1}^{(S)},...,\widetilde{X}%
_{N}^{(1)},...,,\widetilde{X}_{N}^{(S)})}^{-1}(F),\text{ \  \ }F\subseteq
\Omega ,  \label{9}
\end{equation}%
where 
\begin{equation}
\Omega =\mathrm{sp}X_{1}^{(1)}\times \cdots \times \mathrm{sp}%
X_{1}^{(S)}\times \cdots \times \mathrm{sp}X_{N}^{(1)}\times \cdots \times 
\mathrm{sp}X_{N}^{(S)}  \label{9'}
\end{equation}%
and slightly modifying the resulting distribution, we derive for this
correlation $S\times \cdots \times S$-setting scenario on a state $\rho
_{d,N}$ the following LqHV model:%
\begin{eqnarray}
&&\mathrm{tr}[\rho _{d,N}\{ \mathrm{P}_{X_{1}^{(s_{1})}}(B_{1}^{(s_{1})})%
\otimes \cdots \otimes \mathrm{P}_{X_{N}^{(s_{N})}}(B_{N}^{(s_{N})})\}]
\label{10} \\
&=&\dsum \limits_{\omega \in \Omega }\text{ }{\large (}\dprod%
\limits_{n=1,...,N}\chi _{B_{n}^{(s_{n})}}(x_{n}^{(s_{n})}){\large )}\text{ }%
\nu _{S\times \cdots \times S}^{(\rho _{N})}(\omega \text{ }%
|X_{1}^{(1)},...,X_{1}^{(S)},...,X_{N}^{(1)},...,X_{N}^{(S)}),  \notag \\
B_{n}^{(s_{n})} &\subseteq &\mathrm{sp}X_{n}^{(s_{n})},\  \ s_{n}=1,...,S, 
\notag
\end{eqnarray}%
where $\omega =(x_{1}^{(1)},...,x_{1}^{(S)},...,x_{N}^{(1)},...,x_{N}^{(S)})$
and the normalized real-valued distribution $\nu _{S\times \cdots \times
S}^{(\rho _{d,N})}$ is specified in Appendix via the $N$-partite
generalization (\ref{41}) of the bipartite distribution (\ref{26}).

Let us now use the LqHV model (\ref{10}) for finding under von Neumann
measurements at each site of\ the maximal violation $\Upsilon _{S\times
\cdots \times S}^{(\rho _{d,N})}$ by a state $\rho _{d,N}$ of general $%
S\times \cdots \times S$-setting Bell inequalities. This parameter is
defined by relation (51) in \cite{12}.

From Eqs. (40)-(42) in \cite{12} it follows that, in the LqHV terms, the
maximal Bell violation $\Upsilon _{S\times \cdots \times S}^{(\rho _{d,N})}$
takes the form:

\begin{equation}
\Upsilon _{S\times \cdots \times S}^{(\rho _{d,N})}=\sup_{\substack{ %
X_{n}^{(s)},\text{ }s=1,...,S,  \\ n=1,...,N}}\inf \left \Vert \tau
_{S\times \cdots \times S}^{(\rho _{d,N})}(\cdot
|X_{1}^{(1)},...,X_{1}^{(S)},...,X_{N}^{(1)},...,X_{N}^{(S)})\right \Vert
_{var},  \label{11}
\end{equation}%
where: \textrm{(a)} $\tau _{S\times \cdots \times S}^{(\rho _{d,N})}$ $%
(\cdot |X_{1}^{(1)},...,X_{1}^{(S)},...,X_{N}^{(1)},...,X_{N}^{(S)})$ is a
real-valued measure in a LqHV model for the correlation scenario on a state $%
\rho _{N}$ where $S$ qudit observables $X_{n}^{(s)},s=1,...,S,$ are
projectively measured at each $n$-th site; \textrm{(b)} $\left \Vert \tau
_{S\times \cdots \times S}^{(\rho _{d,N})}\right \Vert _{var}$ is the total
variation norm\footnote{%
On this notion, see \cite{dunford} and also Sec. 3\ in \cite{12}.} of a
measure $\tau _{S\times \cdots \times S}^{(\rho _{d,N})};$ \textrm{(c)}
infimum is taken over all possible LqHV models for this scenario and
supremum -- over all collections $X_{n}^{(s)},$ $s=1,...,S$ of qudit
observables measured at each $n$-th site.

From (\ref{11}) and the LqHV model (\ref{10}) it follows 
\begin{equation}
\Upsilon _{S\times \cdots \times S}^{(\rho _{d,N})}\leq \sup_{\substack{ %
X_{n}^{(s)},\text{ }s=1,...,S,  \\ n=1,...,N}}\left \Vert \nu _{S\times
\cdots \times S}^{(\rho _{d,N})}(\cdot \text{ }%
|X_{1}^{(1)},...,X_{1}^{(S)},...,X_{N}^{(1)},...,X_{N}^{(S)})\right \Vert
_{var},  \label{12}
\end{equation}%
where%
\begin{eqnarray}
&&\left \Vert \nu _{S\times \cdots \times S}^{(\rho _{d,N})}(\cdot
|X_{1}^{(1)},...,X_{1}^{(S)},...,X_{N}^{(1)},...,X_{N}^{(S)})\right \Vert
_{var}  \label{13} \\
&=&\dsum \limits_{\omega \in \Omega }\left \vert \nu _{S\times \cdots \times
S}^{(\rho _{d,N})}(\omega \text{ }%
|X_{1}^{(1)},...,X_{1}^{(S)},...,X_{N}^{(1)},...,X_{N}^{(S)})\right \vert 
\notag
\end{eqnarray}%
is the total variation norm of the real-valued measure $\nu _{S\times \cdots
\times S}^{(\rho _{d,N})}$ standing in (\ref{10}). By relation (\ref{42}),
for all states $\rho _{d,N},$ this norm is upper bounded as 
\begin{eqnarray}
\left \Vert \nu _{2\times \cdots \times 2}^{(\rho _{d,N})}(\cdot \text{ }%
|X_{1}^{(1)},X_{1}^{(2)},...,X_{N}^{(1)},X_{N}^{(2)})\right \Vert _{var}
&\leq &d^{\frac{N-1}{2}},\text{ \  \  \  \ for }S=2,  \label{14} \\
\left \Vert \nu _{S\times \cdots \times S}^{(\rho _{d,N})}(\cdot \text{ }%
|X_{1}^{(1)},...,X_{1}^{(S)},...,X_{N}^{(1)},...,X_{N}^{(S)})\right \Vert
_{var} &\leq &d^{\frac{S(N-1)}{2}},\text{ \  \ for }S\geq 3.  \notag
\end{eqnarray}

This and relation (\ref{12}) imply that, for $N$-partite joint projective
measurements on an $N$-qudit state $\rho _{d,N}$, the maximal\emph{\ }%
violation $\Upsilon _{S\times \cdots \times S}^{(\rho _{d,N})}$ by a state $%
\rho _{d,N}$ of general Bell inequalities satisfies the relations%
\begin{eqnarray}
\Upsilon _{2\times \cdots \times 2}^{(\rho _{d,N})} &\leq &d^{\frac{N-1}{2}},%
\text{ \  \  \  \ for \ }S=2,  \label{15} \\
\Upsilon _{S\times \cdots \times S}^{(\rho _{d,N})} &\leq &d^{\frac{S(N-1)}{2%
}},\text{ \  \ for \ }S\geq 3,  \notag
\end{eqnarray}%
for all $N$ and $d.$

Relation (\ref{15}) and our general upper bound (62) in \cite{12} imply
that, under projective parties' measurements at all sites, the maximal
violation by an $N$-qudit state $\rho _{d,N}$ of general Bell inequalities
for $S$ settings per site are upper bounded as 
\begin{eqnarray}
\Upsilon _{2\times \cdots \times 2}^{(\rho _{d,N})} &\leq &\min \{d^{\frac{%
N-1}{2}},\text{ }3^{N-1}\},\text{ \  \  \  \ for \ }S=2,  \label{16} \\
&&  \notag \\
\Upsilon _{S\times \cdots \times S}^{(\rho _{d,N})} &\leq &\min \{d^{\frac{%
S(N-1)}{2}},\text{ }(2S-1)^{N-1},\text{ }(2d)^{N-1}-2^{N-1}+1\},\text{ \  \  \
for \ }S\geq 3,  \notag
\end{eqnarray}%
with%
\begin{eqnarray}
\sup_{d}\Upsilon _{S\times \cdots \times S}^{(\rho _{d,N})} &\leq
&(2S-1)^{N-1},  \label{16'} \\
\sup_{S}\Upsilon _{S\times \cdots \times S}^{(\rho _{d,N})} &\leq
&(2d)^{N-1}-2^{N-1}+1.  \notag
\end{eqnarray}

For $S=d=2$ and an arbitrary $N,$ the upper bound (\ref{16}) takes the form%
\begin{equation}
\Upsilon _{2\times \cdots \times 2}^{(\rho _{2,N})}\leq 2^{\frac{N-1}{2}},
\end{equation}%
that is, reduces exactly to the upper bound known \cite{wolf} for quantum
violations of $2\times \cdots \times 2$-setting Bell inequalities on full
correlation functions in a dichotomic case. This bound is attained \cite%
{wolf} on the Mermin-Klyshko inequality\footnote{%
On this inequality, see \cite{wolf} and references therein and also Sec. 3.3
in \cite{11}.} by the generalized Greenberger--Horne--Zeilinger state (GZH).

Therefore, from the new upper bound (\ref{16}) it follows that, under $N$%
-partite joint von Neumann measurements on an $N$-qubit state, the
Mermin-Klyshko inequality gives the maximal violation not only among all
correlation $2\times \cdots \times 2$-setting Bell inequalities, as it was
proved in \cite{wolf}, but also among all $2\times \cdots \times 2$-setting
Bell inequalities of any type, either for correlation functions, full or
restricted, or for joint probabilities or for both.

The new upper bound (\ref{16}) on the maximal quantum violation of general $%
N $-partite Bell inequalities incorporates and improves our general $N$%
-partite upper bounds (62) in \cite{12} and (19) in \cite{found}.

In the following section, we also explicitly demonstrate that, for all $N,$ $%
S$ and $d,$ our upper bounds in \cite{12, found} and the new upper bound (%
\ref{16}) improve all the upper bounds on the maximal quantum violation of
general Bell inequalities reported in the literature by other authors \cite%
{kaplan, carlos, carlos2, carlos3}.

\section{Discussion}

As it is well-known, in a bipartite case, quantum violations of correlation
Bell inequalities cannot exceed \cite{tsl} the real Grothendieck's constant%
\footnote{%
The exact value of this constant is not known.} $K_{G}^{(R)}=\lim_{n%
\rightarrow \infty }K_{G}^{(R)}(n)\in \lbrack 1.676,1.783]$ independently on
a Hilbert space dimension of a bipartite quantum state and a number $S$ of
settings per site.

For the two-qubit singlet state, this upper bound can be more specified in
the sense that violations of correlation Bell inequalities by the singlet
state are upper bounded \cite{acin} by the real Grothendieck's constant%
\textbf{\ }$K_{G}^{(R)}(3)\in \lbrack \sqrt{2},1.5164]$ of order $3$.

However, upper bounds on quantum violations of general Bell inequalities
have been much less investigated even for a bipartite case.

Let us now specify the new upper bound (\ref{16}) on quantum violations of
general Bell inequalities for some particular cases and compare these
results with the corresponding bounds available in the literature.

\subsection{\textbf{Bipartite case}}

For a bipartite case, the upper bound (\ref{16}) reads 
\begin{eqnarray}
\Upsilon _{2\times 2}^{(\rho _{d,2})} &\leq &\min \{ \sqrt{d},\text{ }3\},%
\text{ \  \ for\  \ }S=2,  \label{17} \\
\Upsilon _{S\times S}^{(\rho _{d,2})} &\leq &\min \{d^{\frac{S}{2}},\text{ }%
2\min \{S,d\}-1\},\text{ \  \ for \ }S\geq 3,  \notag
\end{eqnarray}%
implying%
\begin{equation}
\sup_{d}\Upsilon _{S\times S}^{(\rho _{d,2})}\leq 2S-1,\text{ \  \  \  \ }%
\sup_{S}\Upsilon _{S\times S}^{(\rho _{d,2})}\leq 2d-1.  \label{18}
\end{equation}

From (\ref{17}) it follows that, for the two-qubit case $(d=2)$ with two
settings per site, the maximal violation of general bipartite Bell
inequalities is upper bounded as 
\begin{equation}
\Upsilon _{2\times 2}^{(\rho _{2,2})}\leq \sqrt{2}.  \label{19}
\end{equation}%
Therefore, for the case $S=d=2,$ the new bipartite bound (\ref{17}) reduces
exactly to the attainable upper bound known for quantum violations of\emph{\ 
}Bell inequalities\emph{\ }on joint probabilities and correlation functions%
\emph{\ }in a dichotomic case with two settings per site. The latter result
is due to the Tsirelson \cite{tsirelson} upper bound\footnote{%
The Tsirelson upper bound holds for any Hilbert space dimension $d$.} $\sqrt{%
2}$ on quantum violations of the CHSH inequality and the role \cite{fine,
wolf} which the Clauser-Horne (CH) inequality and the
Clauser-Horne-Shimony-Holt (CHSH) inequality play in the case $N=S=d=2.$

The new bounds (\ref{17}) and (\ref{18}) incorporate and essentially improve
our bipartite upper bounds for general Bell inequalities introduced in \cite%
{12} by Eq. (65) and in \cite{found} by Eq. (19) (specified for $N=2$).

Let us now compare our general bipartite bounds with the corresponding
results available in the literature.

For all $S$ and $d$, our general bipartite upper bound (65) in \cite{12}
and, hence, the new general bipartite upper bound (\ref{17}) are essentially
better than \textrm{(i)} the upper bound%
\begin{align}
\Upsilon _{S\times S}^{(\rho _{2,2})}& \leq 2K_{G}^{(R)}+1,\  \  \  \  \  \  \  \  \
\  \  \  \  \text{for \ }d=2,  \label{20} \\
\Upsilon _{S\times S}^{(\rho _{d,2})}& \leq 2d^{2}(K_{G}^{(R)}+1)-1,\text{ \
\ for \ }d\geq 3,  \notag
\end{align}%
on quantum violations of general bipartite Bell inequalities presented by
theorem 2 in \cite{kaplan}; \textrm{(ii)} the approximate upper bounds%
\footnote{%
Here, symbol $\preceq $ means an inequality defined up to an unknown
universal constant.} 
\begin{equation}
\Upsilon _{S\times S}^{(\rho _{d,2})}\preceq \min \{d,S\},\text{ \  \ }%
\Upsilon _{S\times S}^{(\rho _{d,2})}\preceq \frac{d}{\ln d},  \label{21}
\end{equation}%
and the exact upper bound 
\begin{equation}
\Upsilon _{S\times S}^{(\rho _{d,2})}\leq 2d
\end{equation}%
on quantum violations of general bipartite Bell inequalities found in \cite%
{carlos, carlos2, carlos3} via the operator space theory.

\subsection{\textbf{Tripartite case}}

For a tripartite case, the new upper bound (\ref{16}) for general Bell
inequalities takes the form 
\begin{eqnarray}
\Upsilon _{2\times 2\times 2}^{(\rho _{d,3})} &\leq &\min \{d,\text{ }9\},%
\text{ \  \  \  \ for \ }S=2,  \label{22} \\
\Upsilon _{S\times S\times S}^{(\rho _{d,3})} &\leq &\min \{d^{S},\text{ }%
(2S-1)^{2},\text{ }4d^{2}-3\},\text{ \  \ for\  \ }S\geq 3,  \notag
\end{eqnarray}%
implying%
\begin{equation}
\sup_{d}\Upsilon _{S\times S\times S}^{(\rho _{d,3})}\leq (2S-1)^{2},\text{
\  \  \  \ }\sup_{S}\Upsilon _{S\times S}^{(\rho _{d,2})}\leq 4d^{2}-3.
\label{22'}
\end{equation}

For a three-qubit case $(d=2)$, relations (\ref{22}) reduce to%
\begin{eqnarray}
\Upsilon _{2\times 2\times 2}^{(\rho _{2,3})} &\leq &2,\text{ \  \  \ }%
\Upsilon _{3\times 3\times 3}^{(\rho _{2,3})}\text{\ }\leq 8,  \label{24} \\
\Upsilon _{S\times S\times S}^{(\rho _{2,3})} &\leq &13,\text{ \  \ for \ }%
S\geq 4,  \notag
\end{eqnarray}%
where the bound $\Upsilon _{2\times 2\times 2}^{(\rho _{2,3})}\leq 2$ is
attainable \cite{wolf} on the tripartite Mermin-Klyshko inequality.

The new general tripartite upper bound (\ref{22}) incorporates and improves
our tripartite upper bounds introduced in \cite{12} by Eq. (67) and in \cite%
{found} by Eq. (19) specified for $N=3.$

For all $S$ and $d$, our general tripartite upper bound (67) in \cite{12}
and, hence, the new general tripartite upper bound (\ref{22}) are better
than the upper bound%
\begin{equation}
\Upsilon _{S\times S\times S}^{(\rho _{d,3})}\leq 4d^{2}  \label{23}
\end{equation}%
on quantum violations of general Bell inequalities introduced recently in 
\cite{carlos3} via the operator space theory.

For a more narrow class of Bell inequalities arising in three-player XOR
games, the exact upper bound $\min \{K_{G}^{(R)}\sqrt{S},\sqrt{3d}%
(K_{G}^{(C)})^{3/2}\}$ was presented in \cite{31}. Here, $K_{G}^{(C)}$ is
the complex Grothendieck's constant.

\subsection{\textbf{N-partite case}}

As we stressed above in Sec. 3, the new upper bound (\ref{16}) on quantum
violations of general Bell inequalities incorporates and improves our
general $N$-partite upper quantum violation bounds presented by Eq. (62) in 
\cite{12} and Eq. (19) in \cite{found}.\ 

For all $S$ and $d,$ the general upper bound (62) in \cite{12} and, hence,
the new general upper bound (\ref{16}) are essentially better than the
general upper bound 
\begin{equation}
\Upsilon _{S\times \cdots \times S}^{(\rho _{d,N})}\leq (2d)^{(N-1)},
\end{equation}%
presented recently in \cite{carlos3} via the operator space theory.

For the maximal violation by a Schmidt $N$-partite state of Bell
inequalities arising in $N$-player XOR games, the interesting exact upper
bound $2^{\frac{3(N-2)}{2}}K_{G}^{(C)}$, independent on a Hilbert space
dimension $d$, was presented in \cite{BLV}.

\section{Conclusions}

In the present paper, we have explicitly demonstrated the computational
capabilities of a new type of probabilistic modelling of multipartite joint
quantum measurements -- the local quasi hidden variable (LqHV) modelling
which we introduced and developed in \cite{12, 13, new, found}.

\emph{From the conceptual point of view}, the LqHV modelling corresponds
just to \emph{nonclassicality - }one\emph{\ }of the alternatives, which can
explain (see in Introduction) Bell's violation result in \cite{4} but was
disregarded by Bell \cite{3, 5} in favor of nonlocality. The choice of the
"nonclassicality" alternative results in preserving locality but replacement
of "classical realism", embedded into the HV frame via probability measures,
by "quantum realism" expressed via real-valued measures in the LqHV frame.

\emph{From the mathematical point of view}, the local qHV (LqHV) modelling
frame is very fruitful for quantum calculations and is valid \cite{12, new}
for the probabilistic description of every quantum correlation scenario,
moreover, of all $N$-partite joint von Neumann measurements on each $N$%
-partite quantum state. It is specifically this new type of probabilistic
modelling that allowed us to derive the new upper bound (\ref{16}) on the
maximal quantum violation of general Bell inequalities essentially improving
all the upper bounds reported in the literature via other mathematical
frames, in particular, the upper bounds \cite{carlos, carlos2, carlos3}
found via the operator space theory.

For the $N$-qubit case with two observables per site, the new general upper
bound (\ref{16}) reduces exactly to the attainable upper bound known \cite%
{wolf} for quantum violations of $N$-partite correlation $2\times \cdots
\times 2$-setting Bell inequalities in a dichotomic case. This proves that,
under $N$-partite joint von Neumann measurements on an $N$-qubit state, the
Mermin-Klyshko inequality gives the maximal violation not only among all $%
2\times \cdots \times 2$-setting Bell inequalities on correlation functions
but also among $2\times \cdots \times 2$-setting Bell inequalities of any
type, either on correlation functions, full or restricted, or on joint
probabilities or on both.

\subparagraph{\noindent Acknowledgements.}

I am very grateful to Professor A. Khrennikov for valuable discussions.

\section{Appendix}

Let us first consider a bipartite case $(N=2).$ For simplicity of notations,
denote by $X_{s}$, $s=1,...,S,$ observables measured by Alice and by $Y_{s},$
$s=1,...,S$, measured by Bob.

For a bipartite case, the values of the distribution $\nu _{S\times
S}^{(\rho _{d,2})}(\cdot $ $|$ $X_{1},...,X_{S},Y_{1},...,Y_{S}),$ standing
in (\ref{10}), have the form%
\begin{align}
& 2\nu _{S\times S}^{(\rho _{d,2})}(x_{1},...,x_{S},y_{1},...,y_{S}\text{ }|%
\text{ }X_{1},...,X_{S},Y_{1},...,Y_{S})  \label{26} \\
& ={\large \{}\dprod \limits_{s=1,...,S}\alpha _{X_{s}}^{(+)}(x_{s}\text{ }|%
\text{ }y_{1},...,y_{S}){\large \}}\text{ }\mathrm{tr}{\large [}\rho _{d,2}%
\text{ }{\large \{}\mathbb{I}_{\mathbb{C}^{d}}\text{ }\mathbb{\otimes }\text{
}{\large (}\mathrm{P}_{Y_{1}}(y_{1})\cdot ...\cdot \mathrm{P}_{Y_{S}}(y_{S})+%
\mathrm{h.c.}{\large )}^{(+)}{\large \}]}  \notag \\
& -{\large \{}\dprod \limits_{s=1,...,S}\alpha _{X_{s}}^{(-)}(x_{s}\text{ }%
|y_{1},...,y_{S}){\large \}}\text{ }\mathrm{tr}[\rho _{d,2}\text{ }{\large \{%
}\mathbb{I}_{\mathbb{C}^{d}}\text{ }\mathbb{\otimes }\text{ }\mathbb{(}%
\mathrm{P}_{Y_{1}}(y_{1})\cdot ...\cdot \mathrm{P}_{Y_{S}}(y_{S})+\mathrm{%
h.c.})^{(-)}{\large \}]},  \notag
\end{align}%
\medskip where (i) the term $"+$ $\mathrm{h.c.}"$ means the Hermitian
conjugate of the previous operator; (ii) notations $Z^{(\pm )}$ mean the
positive operators satisfying the relation $Z^{(+)}Z^{(-)}=Z^{(-)}Z^{(+)}=0$
and decomposing a self-adjoint operator $Z=Z^{(+)}-Z^{(-)}$; (ii) the
probability distributions $\alpha _{X_{s}}^{(\pm )}(\cdot |y_{1},...,y_{S}),$
$s=1,...,S,$ are defined, in view of the Radon-Nikodym theorem \cite{dunford}%
, by the relation 
\begin{eqnarray}
&&\mathrm{tr}[\rho _{d,2}\{ \mathrm{P}_{X_{s}}(x_{s})\otimes \mathbb{(}%
\mathrm{P}_{Y_{1}}(y_{1})\cdot ...\cdot \mathrm{P}_{Y_{S}}(y_{S})+\mathrm{%
h.c.})^{(\pm )}\}]  \label{27} \\
&=&\alpha _{X_{s}}^{(\pm )}(x_{s}|y_{1},...,y_{S}))\mathrm{tr}\left[ \rho
_{d,2}\left \{ \mathbb{I}_{\mathbb{C}^{d}}\text{ }\mathbb{\otimes }\text{ }%
\mathbb{(}\mathrm{P}_{Y_{1}}(y_{1})\cdot \ldots \cdot \mathrm{P}%
_{Y_{S}}(y_{S})+\mathrm{h.c.})^{(\pm )}\right \} \right] .  \notag
\end{eqnarray}%
From (\ref{27}) and the relation 
\begin{equation}
\left \Vert \nu _{S\times S}^{(\rho _{d,2})}\right \Vert
_{var}=\sum_{x_{1},...,x_{S},y_{1},...,y_{S}}\left \vert \nu _{S\times
S}^{(\rho _{d,2})}(x_{1},...,x_{S},y_{1},...,y_{S}\text{ }%
|X_{1},...,X_{S},Y_{1},...,Y_{S})\right \vert  \label{28}
\end{equation}%
it follows that the total variation norm $\left \Vert \nu _{S\times
S}^{(\rho _{d,2})}\right \Vert _{var}$ of the distribution (\ref{26}) is
upper bounded as%
\begin{equation}
\left \Vert \nu _{S\times S}^{(\rho _{d,2})}\right \Vert _{var}\leq \frac{1}{%
2}\sum_{y_{1},...,y_{S}}\mathrm{tr}\left[ \widetilde{\rho }_{d,2}\text{ }%
{\Large |}\text{ }\mathrm{P}_{Y_{1}}(y_{1})\cdot \ldots \cdot \mathrm{P}%
_{Y_{S}}(y_{S})+\mathrm{h.c.}{\Large |}\right] ,  \label{29}
\end{equation}%
where $\widetilde{\rho }_{d,2}$ is the qudit state at Bob's site which is
reduced from the two-qudit state $\rho _{d,2}$ and 
\begin{eqnarray}
&&\left \vert \mathbb{\ }\mathrm{P}_{Y_{1}}(y_{1})\cdot ...\cdot \mathrm{P}%
_{Y_{S}}(y_{S})+\mathrm{h.c.}\right \vert  \label{30} \\
&=&{\large (}\mathrm{P}_{Y_{1}}(y_{1})\cdot ...\cdot \mathrm{P}%
_{Y_{S}}(y_{S})+\mathrm{h.c}{\large )}^{(+)}+{\large (}\mathrm{P}%
_{Y_{1}}(y_{1})\cdot ...\cdot \mathrm{P}_{Y_{S}}(y_{S})+\mathrm{h.c.}{\large %
)}^{(-)}  \notag
\end{eqnarray}%
is the absolute value operator.

Calculating operator (\ref{30}), we find%
\begin{eqnarray}
&&\sum_{y_{1},...,y_{S}}\left \vert \text{ }\mathrm{P}_{Y_{1}}(y_{1})\cdot
...\cdot \mathrm{P}_{Y_{S}}(y_{S})+\mathrm{h.c.}\right \vert  \label{31} \\
&=&\dsum \limits_{k_{1},...,k_{S}}\left \vert \beta _{k_{1},...,k_{S}}\right
\vert \text{ }{\LARGE \{}|\phi _{Y_{1}}^{(k_{1}}\rangle \langle \phi
_{Y_{1}}^{(k_{1})}|\text{ }+\text{ }|\phi _{Y_{S}}^{(k_{S}}\rangle \langle
\phi _{Y_{S}}^{(k_{S}}|  \notag \\
&&+{\LARGE (}\frac{\alpha _{k_{s}k_{1}}\beta _{(k_{1},...,k_{S})}^{2}}{\left
\vert \beta _{(k_{1},...,k_{S})}\right \vert ^{2}}|\phi
_{Y_{1}}^{(k_{1}}\rangle \langle \phi _{Y_{S}}^{(k_{S}}|\text{ }+\text{ }%
\mathrm{h.c}{\Large )}{\LARGE \}}^{\frac{1}{2}},  \notag
\end{eqnarray}%
where $\phi _{Y_{s}}^{(k_{s})},$ $k_{s}=1,...,d,$ are the orthonormal
eigenvectors of an observable $Y_{s}$ and 
\begin{eqnarray}
\alpha _{k_{s}k_{1}} &=&\langle \phi _{Y_{S}}^{(k_{S}}|\phi
_{Y_{1}}^{(k_{1}}\rangle ,  \label{32} \\
\beta _{k_{1},...,k_{S}} &=&\langle \phi _{Y_{1}}^{(k_{1})}|\phi
_{Y_{2}}^{(k_{2})}\rangle \langle \phi _{Y_{2}}^{(k_{2})}|\phi
_{Y_{3}}^{(k_{3})}\rangle \cdot \ldots \cdot \langle \phi
_{Y_{S-1}}^{(k_{S-1})}|\phi _{Y_{S}}^{(k_{S})}\rangle .  \notag
\end{eqnarray}%
From (\ref{29}) - (\ref{32}) it follows%
\begin{eqnarray}
\left \Vert \nu _{S\times S}^{(\rho _{d,2})}\right \Vert _{var} &\leq &\frac{%
1}{2}\dsum \limits_{k_{1},...,k_{S}}\left \vert \beta
_{k_{1},...,k_{S}}\right \vert \text{ }\mathrm{tr}{\large [}\widetilde{\rho }%
_{d,2}{\LARGE \{}|\phi _{Y_{1}}^{(k_{1})}\rangle \langle \phi
_{Y_{1}}^{(k_{1})}|\text{ }+\text{ }|\phi _{Y_{S}}^{(k_{S})}\rangle \langle
\phi _{Y_{S}}^{(k_{S})}|  \label{33} \\
&&+{\LARGE (}\frac{\alpha _{k_{s}k_{1}}\beta _{(k_{1},...,k_{S})}^{2}}{\left
\vert \beta _{(k_{1},...,k_{S})}\right \vert ^{2}}|\phi
_{Y_{1}}^{(k_{1})}\rangle \langle \phi _{Y_{S}}^{(k_{S})}|\text{ }+\text{ }%
\mathrm{h.c}\text{ }{\LARGE )\}}^{\frac{1}{2}}{\large ]}.  \notag
\end{eqnarray}%
\bigskip

If $S=2$, then $\beta _{k_{1}k_{2}}=\alpha _{k_{2}k_{1}}^{\ast }=\alpha
_{k_{1}k_{2}}$ and 
\begin{eqnarray}
&&|\phi _{Y_{1}}^{(k_{1})}\rangle \langle \phi _{Y_{1}}^{(k_{1})}|\text{ }+%
\text{ }|\phi _{Y_{2}}^{(k_{2})}\rangle \langle \phi _{Y_{2}}^{(k_{2})}|%
\text{ }+\text{ }{\LARGE (}\frac{\alpha _{k_{2}k_{1}}\alpha _{k_{1}k_{2}}^{2}%
}{\left \vert \alpha _{k_{1}k_{2}}\right \vert ^{2}}|\phi
_{Y_{1}}^{(k_{1})}\rangle \langle \phi _{Y_{2}}^{(k_{2})}|\text{ }+\text{ }%
\mathrm{h.c}\text{ }{\LARGE )}  \label{34} \\
&=&{\large (}|\phi _{Y_{1}}^{(k_{1})}\rangle \langle \phi _{Y_{1}}^{(k_{1})}|%
\text{ }+\text{ }|\phi _{Y_{2}}^{(k_{2})}\rangle \langle \phi
_{Y_{2}}^{(k_{2})}|{\large )}^{2}.  \notag
\end{eqnarray}%
Taking into account (\ref{33}), (\ref{34}) and that sums $\dsum
\limits_{k_{2}}\left \vert \alpha _{k_{1}k_{2}}\right \vert $ and $\dsum
\limits_{k_{1}}\left \vert \alpha _{k_{1}k_{2}}\right \vert $ are upper
bounded by $\sqrt{d},$ we derive for $S=2:$%
\begin{eqnarray}
\left \Vert \nu _{2\times 2}^{(\rho _{d,2})}\right \Vert _{var} &\leq &\frac{%
1}{2}\dsum \limits_{k_{1},k_{2}}\left \vert \alpha _{k_{1}k_{2}}\right \vert 
\text{ }\mathrm{tr}[\widetilde{\rho }_{d,2}\{|\phi _{Y_{1}}^{(k_{1}}\rangle
\langle \phi _{Y_{1}}^{(k_{1})}|\text{ }+\text{ }|\phi
_{Y_{2}}^{(k_{2}}\rangle \langle \phi _{Y_{2}}^{(k_{2}}|\}]  \label{35} \\
&\leq &\sqrt{d}.  \notag
\end{eqnarray}

Let now $S\geq 3$. Since $\mathrm{tr}[\rho \sqrt{Z}]\leq \sqrt{\mathrm{tr}%
[\rho Z]}$ for each $\rho $ and every positive operator $Z$, relation (\ref%
{33}) implies%
\begin{eqnarray}
\left \Vert \nu _{S\times S}^{(\rho _{d,2})}\right \Vert _{var} &\leq &\frac{%
1}{2}\dsum \limits_{k_{1},...,k_{S}}\left \vert \beta
_{k_{1},...,k_{S}}\right \vert {\LARGE \{}\mathrm{tr}{\Large [}\widetilde{%
\rho }_{d,2}{\Large (}|\phi _{Y_{1}}^{(k_{1}}\rangle \langle \phi
_{Y_{1}}^{(k_{1})}|\text{ }+\text{ }|\phi _{Y_{S}}^{(k_{S})}\rangle \langle
\phi _{Y_{S}}^{(k_{S})}|{\Large )]} \\
&&+2{\Large |}\langle \phi _{Y_{S}}^{(k_{S})}|\widetilde{\rho }_{d,2}|\phi
_{Y_{1}}^{(k_{1})}\rangle {\Large |}{\LARGE \}}^{\frac{1}{2}}
\end{eqnarray}%
Taking into account that 
\begin{equation}
{\large |}\dsum \limits_{m}\gamma _{m}\xi _{m}{\large |}\leq {\large \{}%
\dsum \limits_{m}|\gamma _{m}|^{2}\dsum \limits_{m}|\xi _{m}|^{2}{\large \}}%
^{\frac{1}{2}}
\end{equation}%
for all $\gamma ,\xi $, and the relations%
\begin{eqnarray}
\dsum \limits_{k_{1},...,k_{S}}\left \vert \beta _{k_{1},...,k_{S}}\right
\vert ^{2} &\leq &d, \\
\dsum \limits_{k_{1},...,k_{S}}\mathrm{tr}{\LARGE [}\widetilde{\rho }_{d,2}%
{\Large (}|\phi _{Y_{1}}^{(k_{1})}\rangle \langle \phi _{Y_{1}}^{(k_{1})}|%
\text{ }+\text{ }|\phi _{Y_{S}}^{(k_{S})}\rangle \langle \phi
_{Y_{S}}^{(k_{S})}|{\Large )]} &\leq &d^{(S-1)},\text{ }  \notag \\
\dsum \limits_{k_{1},...,k_{S}}{\LARGE |}\langle \phi _{Y_{S}}^{(k_{S})}|%
\widetilde{\rho }_{d,2}|\phi _{Y_{1}}^{(k_{1})}\rangle {\LARGE |} &\leq
&d^{(S-1)},  \notag
\end{eqnarray}%
for $S\geq 3,$ we finally derive:%
\begin{equation}
\left \Vert \nu _{S\times S}^{(\rho _{d,2})}\right \Vert _{var}\leq d^{\frac{%
S}{2}}.  \label{38}
\end{equation}

For an $N$-partite case$,$ we use in the LqHV representation (\ref{10}) the
real-valued distribution 
\begin{equation}
\nu _{S\times \cdots \times S}^{(\rho _{d,N})}(\omega \text{ }|\text{ }%
X_{1}^{(1)},...,X_{1}^{(S)},...,X_{N}^{(1)},...,X_{N}^{(S)}),  \label{39}
\end{equation}%
which is quite similar by its construction to distribution (\ref{26}) but
with the replacement of the terms 
\begin{equation}
\mathbb{I}_{\mathbb{C}^{d}}\mathbb{\otimes }\frac{1}{2}{\large (}\mathrm{P}%
_{Y_{1}}(y_{1})\cdot ...\cdot \mathrm{P}_{Y_{S}}(y_{S})+\mathrm{h.c.}{\large %
)}^{(\pm )}  \label{40}
\end{equation}%
by the $N$-partite tensor product terms 
\begin{eqnarray}
&&\mathbb{I}_{\mathbb{C}^{d}}\mathbb{\otimes }{\Large \{}\frac{1}{2}{\large (%
}\mathrm{P}_{X_{1}^{(1)}}(x_{1}^{(1)})\cdot ...\cdot \mathrm{P}%
_{X_{1}^{(S)}}(x_{1}^{(S)})+\mathrm{h.c.}{\large )}  \label{41} \\
&&\otimes \cdots \otimes \frac{1}{2}{\large (}\mathrm{P}%
_{X_{N}^{(1)}}(x_{N}^{(1)})\cdot ...\cdot \mathrm{P}%
_{X_{N}^{(S)}}(x_{N}^{(S)})+\mathrm{h.c.}{\large )}{\LARGE \}}^{(\pm )}. 
\notag
\end{eqnarray}

As a result, we derive%
\begin{eqnarray}
\left \Vert \nu _{2\times \cdots \times 2}^{(\rho _{d,N})}\right \Vert
_{var} &\leq &d^{\frac{N-1}{2}},\text{ \  \  \ for \ }S=2,  \label{42} \\
\left \Vert \nu _{S\times \cdots \times S}^{(\rho _{d,N})}\right \Vert
_{var} &\leq &d^{\frac{S(N-1)}{2}},\text{ \ for \ }S\geq 3.  \notag
\end{eqnarray}

\end{document}